\documentclass[review]{elsarticle}

\pdfoutput=1

\usepackage{hyperref}
\usepackage{mathtools}
\usepackage{graphicx}
\usepackage{subcaption}
\usepackage{CJKutf8}
\usepackage[bottom]{footmisc}
\usepackage{siunitx}

\journal{Proceedings of the 18th International Symposium on Advanced Intelligent Systems (ISIS2017)}

\begin{document}

\begin{frontmatter}

\title{Analysis of Chinese Tourists in Japan by Text Mining of a Hotel Portal Site}

\author[gidai]{Elisa Claire Alem\'an Carre\'on
\corref{mycorrespondingauthor}}
\ead{s153400@stn.nagaokaut.ac.jp}

\author[gidai]{Hirofumi Nonaka}
\ead{nonaka@kjs.nagaokaut.ac.jp}

\author[nagasaki]{Toru Hiraoka}
\ead{hiraoka@sun.ac.jp}

\address[gidai]{Nagaoka University of Technology, Nagaoka, Japan}
\address[nagasaki]{University of Nagasaki, Nagasaki, Japan}

\cortext[mycorrespondingauthor]{Corresponding author}

\begin{abstract}

With an increasingly large number of Chinese tourists in Japan, the hotel industry is in need of an affordable market research tool that does not rely on expensive and time consuming surveys or interviews. Because this problem is real and relevant to the hotel industry in Japan, and otherwise completely unexplored in other studies, we have extracted a list of potential keywords from Chinese reviews of Japanese hotels in the hotel portal site \textit{Ctrip}\footnote{\label{ctrip}Ctrip: \href {www.ctrip.com/}{\path{www.ctrip.com/}}} using a mathematical model to then use them in a sentiment analysis with a machine learning classifier. While most studies that use information collected from the Internet use pre-existing data analysis tools, in our study we designed the mathematical model to have the highest possible performing results in classification, while also exploring on the potential business implications these may have.

\end{abstract}

\begin{keyword}
Text Mining\sep Sentiment Analysis\sep Support Vector Machine\sep Machine Learning\sep Entropy
\end{keyword}

\end{frontmatter}

\section{Introduction}\label{intro}

In Japan, the population of Chinese tourists has increased over the recent years, with a 107.3\% increase from 2014 to 2015 \cite[][]{jnto2017} and so on. This change in a customer base for many industries, mainly the hotel industry, brings forward a need for an affordable and reliable method to study this new market; which has not been explored in other studies before. Surveys and interviews, previously used both in business and other studies, present difficulties and an increase in costs and time, besides the fact that these studies \cite[][]{chang2010, truong2009} do not target Japan or Chinese customers \cite[][]{ma2008}.

Using the available text reviews of Japanese hotels by Chinese customers available in the website \textit{Ctrip}, our new method proposal is to use a mathematical model using Shannon's Entropy, a concept which can determine a word's probabilistic distribution, to determine a list of relevant keywords to be used in the context of classifying positive and negative emotions in a hotel review using a Support Vector Machine.

There are other studies using similar techniques, mostly with English texts \cite[][]{bollen2011, oconnor2010} but not many that apply these knowledges practically, and those that do \cite[][]{he2013} don't analyze the implications fully. The ones that do focus on Chinese sentiment classification \cite[][]{lee2010-a} or its business applications \cite[][]{zhang2011} are scarce, and use formal words that describe emotions that are unlikely to appear in our reviews. This is why we developed our own sentiment analysis model to provide a practical tool for future marketing choices in our study. In contrast to these studies, in our entropy based keyword extraction we are using words used by a specific group of users that are characteristic of documents written in different emotional states, which has a better capacity for more precise results in the sentiment analysis classification by the Support Vector Machine.

\section{Methodology}\label{methodology}

\subsection{Word Segmentation}\label{segmentation}

For a statistical analysis to be made possible for each word, we segmented the collected Chinese texts without spaces into words using the Stanford Word Segmenter \cite[][]{chang2008}.

\subsection{Entropy Based Keyword Extraction}\label{entropy}

In this study, we based the extraction of the keywords that are influenced by the users’ emotional judgement on the calculation of an entropy value for each word. Speaking in Information Theory terms, Shannon’s Entropy is the expected value of the information content in a signal \cite[][]{shannon1948}. Applying this knowledge to the study of words allows us to observe the probability distribution of any given word inside the corpus. For example, a word that keeps reappearing in many different documents will have a high entropy, given that predicting on which document it would appear becomes uncertain. On the contrary, a word that only was used in a single text and not in any other documents in the corpus will be perfectly predictable to only appear in that single document, bearing an entropy of zero. This concept is shown in the figure below.

\begin{figure}[h]
    \centering
    \begin{subfigure}[b]{0.4\linewidth}
        \includegraphics[width=\linewidth]{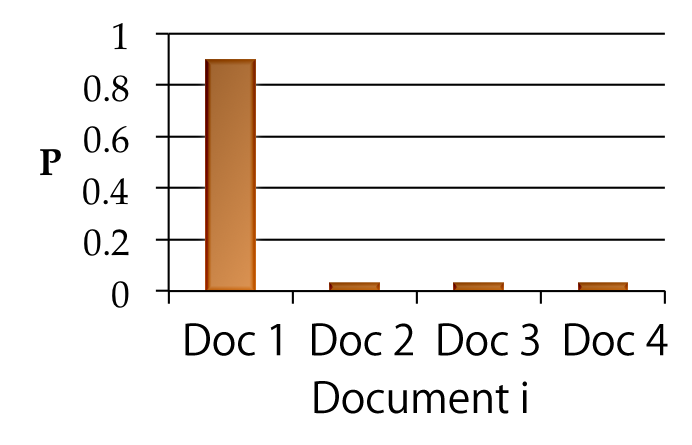}
        \caption{Entropy close to zero.}
    \end{subfigure}
    \begin{subfigure}[b]{0.4\linewidth}
        \includegraphics[width=\linewidth]{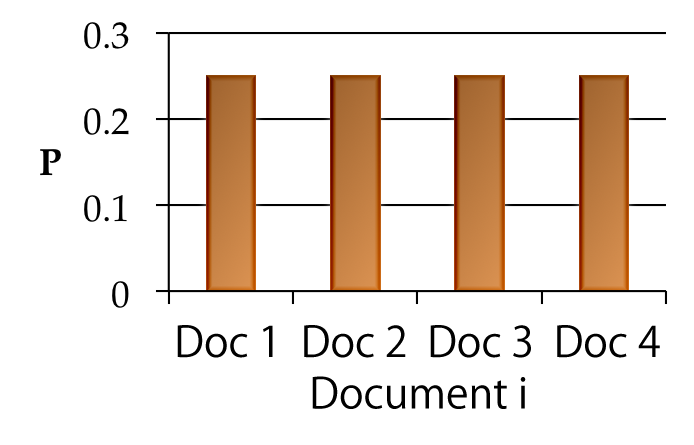}
        \caption{High entropy.}
    \end{subfigure}
\caption{Probabilities of a word \(j\) being contained in a document \(i\).}
\label{fig:entropygraphs}
\end{figure}

With this logic in mind, we used a set of documents that were previously tagged as positive or negative by a group of Chinese students. If a word has a higher entropy in positive documents than in negative documents by a factor of alpha (\(\alpha\)), then it means its probability distribution is more spread in positive texts, meaning that it is commonly used in positive tagged documents compared to negative ones.

To calculate the entropy in a set of documents, for each word \(j\) that appears in each document \(i\), we counted the number of times a word appears in positive comments as \(N_{ijP}\), and the number of times a word appears in negative comments as \(N_{ijN}\). Then, as shown in the formulas below, we calculated the probability of each word appearing in each document shown below as \(P_{ijP}\) (\ref{eq:PijP}) and \(P_{ijN}\) (\ref{eq:PijN}).

\begin{equation}\label{eq:PijP}
P_{ijP} = \frac{N_{ijP}}{\sum_{i=1}^M N_{ijP}}
\end{equation}

\begin{equation}\label{eq:PijN}
P_{ijN} = \frac{N_{ijN}}{\sum_{i=1}^M N_{ijN}}
\end{equation}

We then substitute these values in the formula that defines Shannon’s Entropy. We calculated the entropy for each word \(j\) in relation to positive documents as \(H_{Pj}\) (\ref{eq:Hpj}), and the entropy for each word \(j\) in relation to negative texts as \(H_{Nj}\) (\ref{eq:Hnj}). That is, as is shown in (\ref{eq:lim_Hpj}) and (\ref{eq:lim_Hnj}), all instances of the summation when the probabilities \(P_{ijP}\) or \(P_{ijN}\) are zero and the logarithm of these becomes undefined are substituted as zero into (\ref{eq:Hpj}) and (\ref{eq:Hnj}).

\begin{equation}\label{eq:Hpj}
H_{Pj} = - \sum_{i=1}^M [P_{ijP}\log_2 P_{ijP}]
\end{equation}

\begin{equation}\label{eq:lim_Hpj}
\lim_{P_{ijP}\to0+} P_{ijP}\log_2 P_{ijP} = 0
\end{equation}

\begin{equation}\label{eq:Hnj}
H_{Nj} = - \sum_{i=1}^M [P_{ijN}\log_2 P_{ijN}]
\end{equation}

\begin{equation}\label{eq:lim_Hnj}
\lim_{P_{ijN}\to0+} P_{ijN}\log_2 P_{ijN} = 0
\end{equation}

After calculating the positive and negative entropies for each word, we measured their proportion using the mutually independent coefficients \(\alpha\) for positive keywords and \(\alpha'\) for negative keywords, for which we applied several values experimentally. A positive keyword is determined when (\ref{eq:entropy_pos}) is true, and likewise, a negative keyword is determined when (\ref{eq:entropy_neg}) is true.

\begin{equation}\label{eq:entropy_pos}
H_{Pj} > \alpha H_{Nj}
\end{equation}

\begin{equation}\label{eq:entropy_neg}
H_{Nj} > \alpha' H_{Pj}
\end{equation}

\subsection{Sentiment Analysis Using Support Vector Machine}\label{svm}

In machine learning, Support Vector Machines are supervised learning models commonly used for statistical classification or regression \cite[][]{cortes1995}. Using already classified and labeled data with certain features and characteristics, an SVM learns to classify new unlabeled data by drawing the separating (p-1)-dimensional hyperplane in a p-dimensional space. Each dimensional plane is represented by one of the features that a data point holds. Then each data point holds a position in this multi-dimensional space depending on its features. The separating hyperplane and the supporting vectors divide the multi-dimensional space by minimizing the error of classification. A two-dimensional example is shown below.

\begin{figure}[h]
\centering
\includegraphics[width=20em]{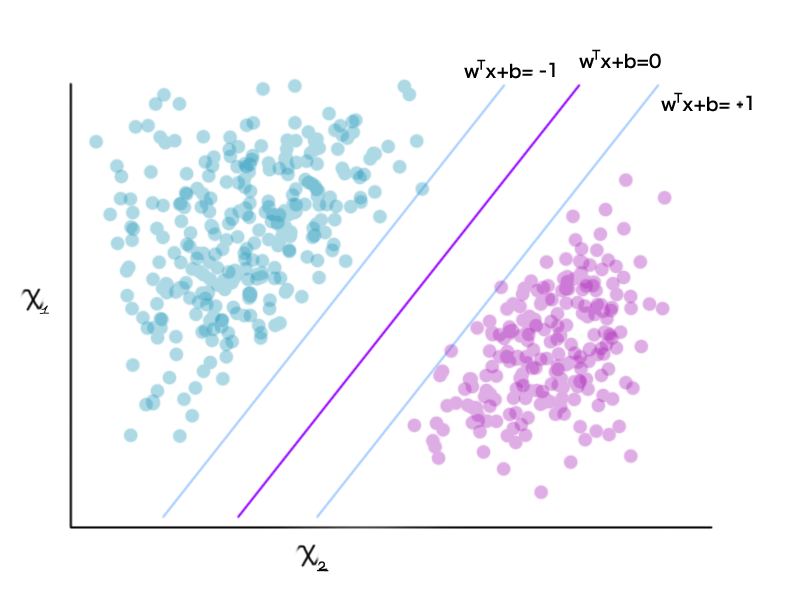}
\caption{Two dimensional example of an SVM classification problem}
\label{fig:svm2d}
\end{figure}

The linear kernel for the SVM classification is defined by the formula (\ref{eq:svm1}) below. The influence that each point of the training data inputs into the vector is defined by their weight \(w_{n}\), included in the Weight Vector \(w\). The bias coefficient \(b\) determines the position of the hyperplane.

\begin{equation}\label{eq:svm1}
f(x) = w^\top x + b
\end{equation}

Then, the conditions shown in (\ref{eq:svm2}) are applied when classifying new data.

\begin{equation}\label{eq:svm2}
f(x) = 
  \begin{cases}
    \geq 0 & y_{i} = +1 \\
    < 0 & y_{i} = -1
  \end{cases}
\end{equation}

Now, the initial condition is set as \(w = 0\). Then each possible separating vector is tested, and when a classification for \(x_{i}\) fails, the value for \(w\) is changed as follows in (\ref{eq:svm4}) by a value of \(\alpha\).

\begin{equation}\label{eq:svm4}
w \leftarrow w + \alpha sign(f(x_i))x_i
\end{equation}

This process is repeated until all points of the training data are classified correctly. The resulting formula for classification (\ref{eq:svm5}) is as follows.

\begin{equation}\label{eq:svm5}
f(x) = \sum_{i=1}^N \alpha_i y_i (x_i^\top \cdot x) + b
\end{equation}

In Natural Language Processing, when it comes to statistically analyzing documents, each possible word in that corpus is a feature, or a dimensional plane. Then, the value of that feature will be marked as the number of times a word \(j\) is contained in a document \(i\).

We implemented this theory in Python using the Support Vector Classifier (SVC) included in the library \textit{scikit-learn}\footnote{\label{scikitlearn}Scikit-Learn \href {http://scikit-learn.org/}{\path{http://scikit-learn.org/}}}. To vectorize the documents into word vector spaces we used the method \textit{CountVectorizer} included in the same library. We then managed these vectors using the mathematics library \textit{numpy}\footnote{\label{numpy}Numpy. \href {http://numpy.org/}{\path{http://numpy.org/}}}.

To evaluate each of our trained machines, we used the K-fold Cross Validation method, which has been proven to provide good results \cite[][]{kohavi1995}. In each test, we calculated then the Precision, Recall, \(F_{1}\) \cite[][]{powers2011} and Accuracy values for our predictions.

\section{Experiments}\label{experiments}

\subsection{Dataset}\label{dataset}

We crawled a total of \num[group-separator={,}]{1541424} html files, from which \num[group-separator={,}]{5938} were unique Japanese hotels. From these pages, we extracted a total of \num[group-separator={,}]{44912} reviews, which turned into \num[group-separator={,}]{286109} separate sentences. In our corpus, there were \num[group-separator={,}]{23443} different words used, from which \num[group-separator={,}]{2802} were noise characters.

\subsection{Sentiment Analysis Experiments}\label{sentiment_experiments}

Before performing the sentiment analysis, we had a group of Chinese student collaborators tag a sample of our collected data, and then calculated the Shannon’s entropy values for each word in positive and negative tags.

We then altered the comparison coefficient \(\alpha\) and \(\alpha'\) to different values, from which we extracted different lists of positive and negative keywords to be tested in an SVM. We changed the values of \(\alpha\) and \(\alpha'\) in 0.25 intervals from 1.0 to 3.75.

In our study, we investigated different parameters to decide the kernel and model of the learning machine. The best performing kernel for the SVM was the Linear one, with a \(C=3.0\).

Based on our K-Fold Cross Validation (\(k=10\)) results, we chose the best performing positive and negative lists. We decided to make a combined list of these two previous lists and train the SVM with this. With this Combined list, the Accuracy was of \(0.90 \pm 0.09\) and the \(F_1\) measure was \(0.93 \pm 0.05\), both excellent results for classification.

\begin{table}[h] \centering
\caption{Results of the K-fold Cross Validation performance tests \protect\footnotemark}\label{tab:svm_f1}
\resizebox{\textwidth}{!}{%
\begin{tabular}{|l|l|l|l|l|}
\hline
\textbf{Keyword List} & \begin{tabular}[c]{@{}l@{}}\textbf{Accuracy} \\ \textbf{Average}\end{tabular} & \begin{tabular}[c]{@{}l@{}}\textbf{Accuracy} \\ \textbf{Std. Dev.} \end{tabular} & \begin{tabular}[c]{@{}l@{}}\(F_1\) \\ \textbf{Average} \end{tabular} & \begin{tabular}[c]{@{}l@{}}\(F_1\) \\ \textbf{Std. Dev.} \end{tabular} \\ \hline
Positive (\(\alpha=2.75\)) & 0.58 & 0.16 & 0.69 & 0.16 \\ \hline
Negative (\(\alpha'=3.75\)) & 0.88 & 0.07 & 0.92 & 0.05 \\ \hline
\underline{\textbf{Combined}} & \underline{\textbf{0.90}} & \underline{\textbf{0.09}} & \underline{\textbf{0.93}} & \underline{\textbf{0.05}} \\ \hline
\end{tabular}%
}
\end{table}

\footnotetext{Not all the lists are displayed. Only the ones with the highest performance rates are shown.}

Following this observation, we predicted whether a sentence was positive or not from the remaining unlabeled data using this model. We then made another prediction of whether a sentence was negative or not using the Negative (\(\alpha = 3.75\) in formula (\ref{eq:entropy_neg})) keyword list trained linear SVM. Having both of these predictions, we decided to classify the documents based on a consensus of the two, having the categories “positive”, “neutral” and “negative”.

\section{Results and Discussion}\label{results_discussion}

\subsection{About the Methodology}\label{disc_method}

We have shown that the entropy based keyword extraction method has many advantages for this field of study. Not only this, but we have obtained a result that exceeds a 0.9 performance (accuracy and \(F_1\), which can be considered as high.

\subsection{About the Results}\label{disc_res}

Now, these are the relevant keywords extracted from the entropy calculations that were used to train the SVM for positive and negative classifying. Because the keyword list for positive words is larger than 100 words, we will only show the top words that are better suited for analysis, such as nouns and adjectives.

\begin{table}[htp]
\centering
\caption{Keywords worthy of note}
\label{tab:keywords}
\resizebox{\textwidth}{!}{%
\begin{tabular}{|l|l|l|l|l|l|}
\hline
\multicolumn{1}{|c|}{\textbf{Word}} & \multicolumn{1}{c|}{\textbf{Translation}} & \multicolumn{1}{c|}{\textbf{Word}} & \multicolumn{1}{c|}{\textbf{Translation}} & \multicolumn{1}{c|}{\textbf{Word}} & \multicolumn{1}{c|}{\textbf{Translation}} \\ \hline
\begin{CJK}{UTF8}{gbsn} 园林 \end{CJK} & garden & \begin{CJK}{UTF8}{gbsn} 烤肉 \end{CJK} & Barbecue & \begin{CJK}{UTF8}{gbsn} 人员 \end{CJK} & personnel \\ \hline
\begin{CJK}{UTF8}{gbsn} 花园 \end{CJK} & flower garden & \begin{CJK}{UTF8}{gbsn} 中文 \end{CJK} & Chinese text & \begin{CJK}{UTF8}{gbsn} 华人 \end{CJK} & Chinese person \\ \hline
\end{tabular}%
}
\end{table}

\section{Conclusion and Future Work}

In our study, with the objective to understand what Chinese customers of Japanese hotels demand and feel we extracted keywords from a Chinese portal site with reviews of Japanese hotels. In the evaluation process of our ma- chine learning experiments, we obtained our highest performance (\(F_{1} = 0.93 \pm 0.05\) and \(Accuracy = 0.90 \pm 0.09\)) using both positive and negative keywords extracted using Shannon’s entropy as a base to train a linear kernel SVM.

In regards to the needs of Chinese customers of Japanese hotels we found that the Chinese customers in our database were less concerned with price and more particularly with food, gardens or parks near the hotel, good service and transport, as well as the use of Chinese language to cater to them specifically.

It could be thought that some of the keywords being used were not about the hotels themselves, but for the surrounding area, which is a subject we will leave for future works. Another subject to investigate is how these keywords could be rated to determine the most important of the demands of Chinese customers of Japanese hotels.

\clearpage

\section*{References}

\bibliography{isis-bib}

\end{document}